\documentclass[runningheads]{llncs}
\usepackage{graphicx}
\usepackage[inline]{enumitem}
\usepackage{cite}

\begin{document}
\title{Medical systems, the role of middleware and survey on middleware design}
%
%
\author{Imad Eddine Touahria\inst{1,2}}

%
%
\institute{Department of Telematics engineering, Universidad Carlos III de Madrid, Leganes, Madrid, Spain \and
Department of computer science, Ferhat Abbas Setif-1 University, Setif, Algeria  \vspace{5px}
\email{100370038@alumnos.uc3m.es } \email{imad.touahria@univ-setif.dz}\\
}

\maketitle              
\begin{abstract}
\label{sec:abstract}
The integration of medical devices in the patient treatment process becomes increasingly important due to the efficiency of the technology. On the one hand, medical devices hardware is more powerful and its integration with the software platforms is improved. These devices are able to transfer accurate data to the clinician and offer the possibility of sharing data with other devices or computational servers for advanced analysis. On the other hand, medical software platforms are appearing to provide advanced functionality on top of these medical hardware devices. 
In this context, the role of the software is essential, not only at the highest abstraction level that provides the application business logic; the role of the underlying connecting software (the communication middleware or distribution software) is essential. These are capable of providing advanced connectivity functions, very efficiently, and within appropriate time bounds.
This paper reviews the state of the art on middleware as facilitator for interconnection among devices and also describes a number of initiatives (such as the Integrated Clinical Environment --ICE) and projects that further extend the underlying distribution software towards the clinical domain as device interconnection facilitators.

\end{abstract}
\section{Introduction \& motivation }
\label{sec:introduction}

By 2025 the number of patients requiring continuous monitoring will approximate 1.2 billion \cite{world2003diet}, this number shows the growing need to medical devices where continuous monitoring includes the use of devices (mobile or bedside) to improve the quality of care and safety.

The integration of technology in health care has decreased levels of mortality and has, overall, improved patient care.
Health monitoring solutions are present everywhere from a small medical application deployed on a PDA (Personal Digital Assistant) to measure heart rate to complex systems deployed on a hospital server that ensures monitoring of a patient with critical or chronic diseases and clinician decision support on diagnosis. Medical devices are an example of the new era of healthcare, as these devices offer new possibilities for physicians and a new perspective of healthcare from a data-centric and efficient  interoperability (objective of this study). 

The use of medical devices is essential in some critical clinical situations or when the patient is at home and needs remote diagnosis, these devices can be mobile \cite{7336385,8363196,8336776} or installed close to the patient clinical bed \cite{5070972}. The hardware/software of each device vary from others and depends on the manufacturer policy, these differences may generate some integration problems in the clinical situation. The hospital network contains a set of computational servers, switches, data recorders and medical devices/equipment; the latter ones have to be integrated in the network as Plug and Play (PnP) devices, in order to achieve this, device manufacturers have to respect standards for medical data exchange. 


Improving the assistance given to patients depends greatly on the capacity of medical systems to collect large data amounts in real-time, exchange it, process it, and create new knowledge that assists medical decisions. Collecting data from multiple sensors and devices can only be done if \textit{interoperability} is achieved effectively, efficiently, and in a timely manner.

Therefore, interoperability in the medical domain is an essential objective, that is not so easy to achieve given the diversity of stakeholders that vary from medical devices to computational units, to human actors. 
\textit{"Medical device interoperability is the ability to safely, securely, and effectively exchange and use information among one or more devices, products, technologies, or systems. This exchanged information can be used in a variety of ways including display, store, interpret, analyze, and automatically act on or control another product"} \cite{fdainteroperability}. OR.NET \cite{ORNET} and ICE \cite{CIMIT}
are examples of projects aiming to realize a safe and interoperable network of medical devices and that are studied in details in this paper. 

The boost of device interconnecting has also reached e-Health, where medical devices are increasingly connected to either other devices and/or to computational servers. To support such interconnection in a \textit{safe} way, medical systems actors convened on the need of a common environment to facilitate interoperability and safety. The Integrated Clinical Environment (ICE) is a new solution that includes many stakeholders where the final goal is to realize an interoperable network of medical device in a \textit{safe} way \cite{8029619}, to achieve this medical devices manufacturers must rely to standards defined by ICE and/or other interoperability projects. Communication standards in medical devices are away to be used by manufacturers even in the same company and for devices that are from the same range, as a result, devices that have the same communication standards may not be able to communicate. According to the west health organization \cite{WEST}, \$30 billion dollars can be saved in the USA by the adoption of functional interoperability for medical devices.

\subsection{Paper structure}
\label{sec:structure}
The paper is structured as follows.  
Section \ref{sec:rolemw} illustrates the importance of the middleware in a distributed medical system.
Section \ref{sec:group} describes the seminal work on middleware design and distribution of the group that provides the baseline for designing middleware specifically tailored to the needs of medical systems that are networked.
Section \ref{sec:standards}  describes medical devices and medical communication standards as well as illustrates some examples of frameworks for medical devices coordination and interoperability projects.
Section \ref{sec:conclusion} 
 draws some conclusions and presents the future work
lines.

\section{The role of the distribution software in medical systems}
\label{sec:rolemw}

The lack of connectivity among medical devices is caused by differences in operating systems, networking ports, data formatting and encoding. Many important projects are trying to realize an effective integration of these devices in the clinical room and achieve an interoperable and secure solutions. These are all lower levels of technicalities that typically are taken care of by the distribution software. 
The most widely adopted standards such as ICE need to build on top of efficient solutions such as DDS \cite{DDS}. 

However, there is a long way on design of the middleware for efficient communications across nodes to provide interoperability, timeliness, flexibility, reconfiguration, and many other characteristics. 

In this paper, a study of a selected set of middleware research works have been analyzed, mostly those related to the research on distributed real-time systems group that are the background to the research on this field introduced there. This broad work goes from resource management techniques for operating systems (that are essential to middleware), middleware design, service oriented middleware, real-time reconfiguration, and hardware accelerated execution of middleware. Now, this work is being extended to target medical systems. Therefore, to complement this survey, a further description of higher abstraction solutions for the medical domain is given and also the applications of the previous work to specific medical scenarios.

\section{Background on distribution sofware design}
\label{sec:group}

This section presents a selected set of works related to the design of real-time middleware and resource management. These contributions are essential to achieve efficient communication and distribution infrastructures that support efficiently the interoperation of distributed medical devices.
\vspace{0.2cm}

\noindent \textbf{Cyber-physical systems}. Medical devices are integrated as part of cyber-physical systems (CPS) that monitor the physical conditions of systems and actuate on them or help in deciding how a human physician will actuate on the patients. Because they are cyber-physical systems, they require rigorous design techniques that tend to be based on formal methods supporting the verification of their properties. 

The evolution and flexible nature of a CPS in the context of medical systems is justified by the fact that they are related to human monitoring and this is a quite unstable situation that can raise different unexpected events. For this reason, supporting evolution and flexibility are among the greatest challenges to be tackled. Reconciling the uncertainty derived from the unknown monitored conditions of a patient, etc., and the need to react in real-time is extremely hard to integrate all together as all the possible situations that can be encountered at run time can not be known a priori. In \cite{JSSPoliMi}, a formal design is described based on Petri nets to model systems that can evolve; this technique was also explored in \cite{Compsac2014}. A different formal method approach is applied in \cite{Bersani18} and \cite{BersaniHASE18}. 

In the above works, the focus is placed on the design of the software component interactions relative to their timing properties and other behavioral parameters that are modeled. However, the communication across the above components is a key aspect that must be analyzed to achieve communication and interaction infrastructures that support timely interaction and variable conditions such as load peaks or coexistence of components with heterogeneous resource usage patterns.
For this purpose, there are also a number of contributions for the design of distribution middleware for CPS as middleware is the key software layer that is capable of abstracting distribution and interaction, masking situations where a node can receive a peak of requests from other nodes; the systems must be resilient to these and other situations, and they have to continue to work at all times. The design of adaptive middleware is provided in OmaCy architecture \cite{omacy}. In \cite{FGCSreview} and \cite{JSAReiser}, an analysis of this problem is outlined. In \cite{CSIcpsmiddleware}, the design of a scheme for attending simultaneous requests is provided. In \cite{RPiDDS}, a model for integrating the Data Distribution Software with single board computers and Raspberry Pi is provided; this is further reworked in \cite{DDSFace} for a different domain such as avionics. Also, there have been a number of dedicated research contributions to building real-time facilities in middleware such as \cite{jsareaction2012, JSASI2017, JSAReiser}, among others; or building abstractions for utilization of multiple interaction paradigms such as \cite{adaeurope11} or \cite{adaeurope12}.

\vspace{0.2cm}

\noindent \textbf{Medical systems}. The role of middleware in medical systems is key to achieve safe execution and timely operation that can, in the end, save lives. Well proven service-oriented architectures such as iLAND \cite{ilandTII} have been integrated with ICE in \cite{sensorsice}. The reconfiguration capabilities \cite{isorcyork} and timely communication capacities of iLAND have been proposed to be the core interoperation backbone for ICE. A number of studies for profiling the actual performance of communication middleware such as \cite{uc3muma} has been particularized for medical systems in a number of works such as \cite{sigbediland} for the Internet Communications Engine and \cite{sigbedamqp} for AMQP. Moreover, a number of improvements to their execution by making the middleware aware of the underlying hardware structure have been undertaken in \cite{sac2017} and the benefits of this acceleration in medical systems for remote patient monitoring has been exemplified in \cite{uc3mupv}.

\vspace{0.2cm}


\noindent \textbf{Resource management and components for real-time}. This section continues to move down the abstraction level in the way towards designing flexible cyber-physical systems focusing at the middleware view.

The basics of the middleware design is the control of the execution at thread or task level. There are a number of contributions on the architecting of real-time middleware from a real-time perspective such as the following such as Hola-QoS \cite{HolaQOS}; using real-time scheduling for distributed actions \cite{ares}; real-time quality of service management \cite{ambient}; mode changing policies for timely execution \cite{modechanges,fcgs12}; agents modeling real-time properties \cite{mata}; identification of Linux kernel properties for improving locking \cite{peteradaeurope}; architecting open source projects for Linux \cite{peterspe}; analysis of temporal behavior at bus level within a multiprocessor \cite{groba}; reconfiguration scheduling \cite{aina}.

Other contributions for service oriented systems have been \cite{soca}; or component-based modeling over QoS networking \cite{demiguel}; have progressively shed light over how to handle execution in systems using more abstract modeling like encapsulation through services and separating application from networking responsibilities.



\section{Medical devices and medical communication standards}
\label{sec:standards}

The world health organization defines a medical device \cite{WHODEVICE} as:
``\textit{
Any instrument, apparatus, implement, machine, appliance, implant, reagent for in vitro use, software, material or other similar or related article, intended by the manufacturer to be used, alone or in combination, for human beings, for one or more of the specific medical purpose(s) of:} 
\begin{enumerate*}
  \item \textit{Diagnosis, prevention, monitoring, treatment or alleviation of disease}.
  \item \textit{Diagnosis, monitoring, treatment, alleviation of or compensation for an injury.}
  \item \textit{Investigation, replacement, modification, or support of the anatomy or of a physiological process.}
  \item \textit{Supporting or sustaining life.}
  \item \textit{Control of conception.}
  \item \textit{Disinfection of medical devices.}
  \item \textit{Providing information by means of in vitro examination of specimens derived from the human body.}
\end{enumerate*}
''

\begin{figure}[h!]
\centering
\centerline{\includegraphics[scale=0.8]{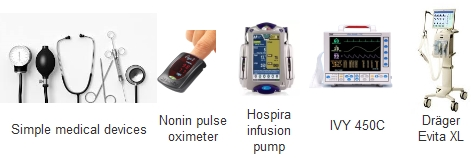}}
\caption{Examples of medical devices}
\label{fig:md}
\end{figure}

As shown in figure \ref{fig:md} and according to the previous definitions, a wide variety of devices can be considered as a \textit{medical device}. Medical devices that have software/hardware components, measure patient vital signs, have network or data sharing ports and are even mobile or bedside are included in this study.

\subsection{Software and hardware characteristics of medical devices}
\label{sec:characteristics}

The strict regulations on design and development of medical systems are a challenge. FDA (Food and Drug Administration) \cite{FDA} is considered as the most important organization that establishes regulations for medical devices development. There have been some critics on its role in minimizing innovation, e.g., X-Ray machines innovation caused by strict FDA strict regulations \cite{ekelman1988technological}. The European Union adopted the MDD (Medical Device Directive) that is a set of standards and regulations that manufacturers of medical devices must respect in order to set up there devices.

Among these regulations there are: the Medical Device Directive (MDD 93/42/EEC), the In Vitro Diagnostic Medical Device Directive (IVDMDD98/79/EC) and the Active Implantable Medical Device Directive (AIMDD 90/385/EEC) \cite{french2012medical} despite these different regulation development, the countries involved most in medical device regulation established the Global
Harmonization Task Force (GHTF) and, after that, the International Medical Device Regulators Forum (IMDRF) \cite{IMDRF} appeared, it objective is to accelerate international medical device regulatory harmonization and convergence with respect to safety, performance and quality of medical devices.

According to WHO (World Health Organization) \cite{WHODEVICE} definition, a medical device can be a software, and IMDRF \cite{IMDRF} defined the concept of \textit{``Software as a Medical Device''} that is a software intended to be used for one or more medical functions that perform them without being part of a hardware medical device.  But, according to the view of this study on medical devices, a software is a medical device when the functional properties of the software are enough to handle the situation where the software is used as a medical device. 

Software is progressively playing an increasingly important role in healthcare \cite{6334808}, and consequently, so is the role of the middleware that it contains. A medical device software has safety requirements, and, therefore, it must operate adhering strictly to parameters of accuracy, high integrity, security, and should be verified and validated through software development methodologies. This is in line with the requirements of cyber-physical systems.

The main objective to fix when developing medical device software is \textit{safety} \cite{7814546}, in order to achieve safety, the software of medical devices has to comply to standards like: European MDD \cite{199342EC, 200747EC}, ISO 13485 \cite{134852012}, IEC 62366 \cite{623662007}, ISO 14971 \cite{149712012} and others.

The hardware characteristics of medical devices vary from a manufacturer to another The weight, size, ports, network connections, displays, mobile/bedside and cable connections can vary depending on the use of the device and the clinical situation where it is deployed; so that different models for a single manufacturer are also common. Even the materials used to construct the medical devices must be analyzed to avoid electrical shocks or allergy of the patients. This study concentrates on input/output ports and network connections of the device. 

\begin{figure}[h!]
\centering
\centerline{\includegraphics[scale=0.5]{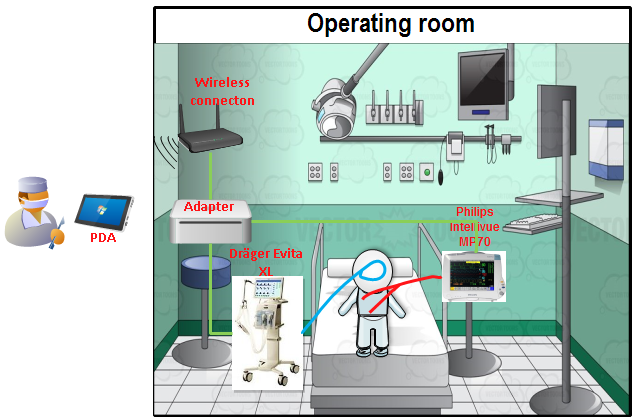}}
\caption{Example of medical devices use in a clinical situation}
\label{fig:clinical}
\end{figure}

In order to communicate medical devices, these  are provided with input/output ports, data output ports vary across suppliers of medical devices. The most common ports are RS 232 port (DB-9, DB-15, and DB-25),
RJ 45, wireless LAN, Bluetooth, USB, or some proprietary data connection systems
developed by suppliers for using data by their own IT systems. The following are the most common hardware connections used by suppliers to input data into the device: PS/2 (for
supporting  keyboard or a mouse inputs),
USB, RS 232, and digital data input \cite{8029619}. Figure \ref{fig:clinical} shows how medical devices are integrated in a clinical situation and the way how medical devices are connected to the patient body, to each other and to networking devices.

\subsection{Networking and messaging.}
\label{sec:networking}

Communication and data exchange between medical devices is a part of \textit{interoperability}. Data and communication exchange between medical devices is characterized by a set of standards that are detailed in what follows.

\noindent \textbf{DICOM} (Digital Imaging and Communications in Medicine) is a standard developed by the National Electrical Manufacturers
Association (NEMA) to store retrieve and transfer medical imaging from various medical devices such as scanners, printers, network hardware etc \cite{7988050}. DICOM promotes digital information exchange between medical imaging equipment and
other systems, and therefore medical imaging is done between physicians or medical centers in an interoperable way.

\noindent \textbf{HL7 version 2}  is a standard of data exchange in healthcare, designed to support data messaging in a centralized or distributed environment and propose interfaces communication to stakeholders that do not adhere to data exchange standards. 95\% of hospitals, 95\% of medical related equipment and information systems in the whole America use HL7 standard; also, it is used in Germany, Japan and other developed countries \cite{5974909}. 

\noindent \textbf{HL7 version 3} is a newer version of HL7 version 2 that uses the eXtensible Markup Language (XML) as a powerful tool in the web to transfer data. Also, version 3 integrates web services and the Web Services Description Language (WSDL) \cite{5687707}. Version 3 of HL7 promotes semantic interoperability defined a more explicit methodology for the development of messages \cite{7377675}.

\subsection{Interoperability.}
\label{sec:interoperability2}

Integrated in the same clinical situation, medical devices must communicate in order to provide high quality healthcare. Interoperability is the basis of data exchange in the clinical environment, 
When connected to a PC, the printer doesn't need to be programmed to be able to communicate with the PC, interoperability between medical devices is expected to achieve that level.

\noindent \textbf{HL7}. The main objective of HL7 is to achieve an interoperable network of medical devices in a secure way. For this, it deals with data and the way it is transferred between devices and HIS (Health Information Systems). Also, other technologies cited above can be classified under more than one section.

\noindent \textbf{ISO/IEEE 11073} is a family of standards for plug and play interoperability of medical devices, defines a common framework for the establishment of a unified data structure model \cite{7494019}. ISO/IEEE 11073 (X73PHD)\cite{11073,staff2010iso} objective is to standardize Personal Health Devices (PHDs) and allow semantic interoperability of medical devices by defining the structure of data and the protocol for information delivery between individual
medical devices (Glucose meter, Weight scale, Blood pressure
monitor, etc.) and the manager (computer, smartphone, set top box, etc.), which collects and manages the information from the individual medical devices \cite{6152915}. 

\noindent \textbf{FHIR} (Fast Healthcare Interoperability Resources) of HL7 is the most recent standard of the series of standards (HL7 v2, HL7 v3, CDA) developed by HL7 \cite{6627810}, it is a framework for the exchange of electronic health
records (EHR) data that combines the evolving modern technologies and market needs \cite{8031128}. FHIR aims to follow the Representational State Transfer (REST) architectural style as presented by Fielding \cite{Fielding:2000:ASD:932295} and presents the stakeholders from the healthcare as resources: medical devices, clinicians, medications, IT structures, etc.

\noindent \textbf{DPWS} (Devices Profile for Web Services) uses SOA (Service Oriented Architecture) for providing interoperable, cross-platform, cross-domain, and network-agnostic access to devices and their services \cite{7325198}. DPWS is used for embedded devices with limited resources by enabling Web services using IoT applications. DPWS requires WSDL (Web Service Description Language) and SOAP (Simple Object Access Protocol) to communicate the device services, but it does not need a registry like UDDI (Universal Description, Discovery and Integration) for services discovery. DPWS aims to achieve interoperability by using the loosely coupled concept of Web services over the MD operation and data encryption.

\noindent \textbf{MDPWS} (Medical Devices Profile for Web Services) is a part of IEEE 11073-20702 series of standards and uses the principles of DPWS but for medical devices interoperability domain with some modifications like the restricted security mechanisms of MDPWS comparing with DPWS, e.g. the usage of client authentication with HTTP authentication is withdrawn in favor of using X.509.v3 certificates \cite{7390446}. MDPWS uses the principles of DPWS with respect to the high acuity patient environment and the complexity of medical devices.

\subsection{Coordination frameworks}

\subsubsection{ICE}
\label{sec:ICE}

(Integrated Clinical Environment) \cite{CIMIT} architecture was defined in 2009 in ASTM (American Society for Testing and Materials) F2761 standard. ICE aims to cover the heterogeneity of medical devices by implementing a plug and play environment of medical devices and creating a communication gateway between them, where messages and commands are exchanged successfully. 
Suppliers of medical devices must adjust the set of specifications provided by ICE in order to create devices that are ICE-compliant: medical devices must have a network output port and must produce data that can be managed through ICE interfaces. ICE aims to:

\begin{list}{$\bullet$}{\leftmargin=1em \itemindent=0em}
\item Improve patient safety by coordinating medical devices actions and avoid incorrect medical decisions generated by a faulty device operation.
\item Ensure support for clinicians in their monitoring and treatment operation, where clinical aid information is generated by a set of workflows implemented in the ICE framework logic.
\item Create a flexible communication bus between medical devices, servers running medical applications, and the clinicians.
\item Implement an interoperable network of medical devices and computational servers where data and messages are exchanged in real time.
\item Define standards for the hardware and software characteristics or dimensions of medical devices that will be used by manufacturers to produce  medical devices that comply with ICE.
\end{list}

 In ICE-based systems \textit{safety} is the ability to implement interoperability between heterogeneous medical devices in a single high acuity patient environment where communication is done via software or hardware interfaces. ICE aims to improve patient \textit{safety} by elaborating and deploying interoperability of the medical devices, thus, creating an interoperate communication bus between heterogeneous medical devices where messages and commands are exchanged. ICE defines the following objectives in order to underline and improve patient \textit{safety}:

\begin{list}{$\bullet$}{\leftmargin=1em \itemindent=0em}
\item Encapsulates errors generated by medical devices that affect patient safety. Errors can be wrong values for a given vital sign or a false alarm.
\item Records errors in order to achieve a system that can predict a given behavior when the system faces pre-saved errors.
\item Creates a support workflow for clinicians to notify them if the patient health status is deteriorating.
\item Introduces a risk management process and defines risk levels that are directly related to the policy of the manufacturer.
\item Defines the concept of \textit{basic safety} as the elimination of medical risks directly related with the physical hazards when a medical device is used in normal conditions \cite{goldman2008medical}. Safety is a wider definition of basic safety. 
\item Supports alarms generated by medical devices. Each alarm is provided to the clinician with its cause and time of generation.

\end{list}



\subsubsection{OR.NET.}
\label{sec:OR.NET}
OR.NET is a solution developed by German academics and industrials for medical devices integration and medical systems interoperability in the operating room and it surroundings. The objective of OR.NET is to develop basic concepts for the secure dynamic networking of computer-controlled medical devices in the operating room and clinic  \cite{ORNET}.  In the end, these concepts are evaluated and transformed to standards. OR.NET aims to create a service-oriented architecture for the safe and secure dynamic interconnection of medical devices in the OR context \cite{7318708}. OR.NET project aims to:

\begin{list}{$\bullet$}{\leftmargin=1em \itemindent=0em}
\item Develop standards for dynamic integration of medical devices in the OR and beyond.
\item Solve the problem of the approval of modular devices in an open networked OR system.
\item Define risk management and usability analysis processes via open interfaces in all the OR system.
\item Achieve hard real time communication between the medical devices using SRTB (Surgical Real Time
Bus).
\item Define an architecture that has security and privacy as first class citizens in order to protect data and patient safety.
\item User interface profiles analysis including GUI interaction elements and input/output ports to develop  reliable and usable medical devices.
\item Adher an use IEEE 11073 family standards including 20701, 20702 and 10207.

\end{list}

Furthermore, requirements were identified for test scenarios to
verify that the devices safely communicate with each other. This
includes the validation of the standard conformity of the messages
being transferred and of the way the systems behave on receiving
regular messages or also in various exceptional situations,
such as network problems, dealing with faulty data, or how to react
when unauthorized users try to take over control.

OR.NET also allows the use of different communication
protocols (e.g. DICOM and HL7 Version 2). OSCP explicitly
does not try to replace these widespread protocols. Instead, dedicated
gateways are specified that enable the operation of the
DICOM and HL7 protocols despite the separation of OR network
and the hospital network.


\section{Conclusion and future works}
\label{sec:conclusion}

This paper has summarized the importance of distribution and communication middleware inside distributed medical systems that are composed of medical devices that can be highly interconnected. The interaction across the elements of a distributed medical system is as efficient as the underlying middleware software that supports its interoperation, coordination, and communication. For this purpose, there are a number of contributions at the levels of entity (component, service, etc.) definition, entity interoperability, and resource management that are relevant in this context. This paper has given value to the work of the distributed real-time systems lab in this domain as a baseline for building efficient middleware specifically tailored to distributed medical systems.

\label{Bibliography}
\bibliographystyle{plain}
\bibliography{References}

\begin{thebibliography}{10}

\bibitem{CIMIT}
{CIMIT} - {Center for Integration of Medicine and Innovative Technology}.
\newblock \url{http://www.cimit.org/}.
\newblock Accessed: 2018-07-02.

\bibitem{FDA}
{FDA} - {US {F}ood and {D}rug Administration}.
\newblock \url{https://www.fda.gov/}.
\newblock Accessed: 2018-07-13.

\bibitem{11073}
Ieee 11073 {P}ersonal health devices.
\newblock \url{ http://www.11073.org/}.
\newblock Accessed: 2018-08-16.

\bibitem{IMDRF}
International {M}edical {D}evice {R}egulators {F}orum.
\newblock \url{http://www.imdrf.org/}.
\newblock Accessed: 2018-07-13.

\bibitem{ORNET}
{OR.NET} - {Operating Room project}.
\newblock \url{http://ornet.org/}.
\newblock Accessed: 2018-07-01.

\bibitem{WEST}
West health organization.
\newblock \url{http://www.who.int/en/}.
\newblock Accessed: 2018-07-12.

\bibitem{WHODEVICE}
World {H}ealth {O}rganization - {M}edical {D}evice {F}ull {D}efinition.
\newblock \url{http://www.who.int/medical-devices/full-deffinition/en/}.
\newblock Accessed: 2018-07-12.

\bibitem{199342EC}
European {C}ouncil - {C}ouncil {D}irective 1993/42/{E}{C}.
\newblock ed.{L}uxembourg: {O}fficial {J}ournal of the {E}uropean {U}nion,
  1993.

\bibitem{200747EC}
European {C}ouncil - {C}ouncil {D}irective 2007/47/{E}{C}.
\newblock ed.{L}uxembourg: {O}fficial {J}ournal of the {E}uropean {U}nion,
  2007.

\bibitem{623662007}
International {E}lectrotechnical {C}ommission - {I}{E}{C} 62366:2007.
\newblock Medical devices - {A}pplication of usability engineering to medical
  devices, ed. Geneva,, 2007.

\bibitem{149712012}
International {O}rganization of {S}tandardization - {I}{S}{O} 13485:2012.
\newblock Medical devices - {A}pplication of risk management to medical
  devices, ed. Geneva, 2012.

\bibitem{134852012}
International {O}rganization of {S}tandardization - {I}{S}{O} 13485:2012
  {M}edical {D}evices, {Q}uality management system.
\newblock Requirements fo regulatory purposes, ed. {G}eneva, 2012.

\bibitem{ares}
Alejandro Alonso, Marisol Garc{\'{\i}}a{-}Valls, and Juan~Antonio de~la Puente.
\newblock Assessment of timing properties of family products.
\newblock In {\em Development and Evolution of Software Architectures for
  Product Families, Second International {ESPRIT} {ARES} Workshop, Las Palmas
  de Gran Canaria, Spain, February 26-27, 1998, Proceedings}, pages 161--169,
  1998.

\bibitem{soca}
Gaetano~F. Anastasi, Tommaso Cucinotta, Giuseppe Lipari, and Marisol
  Garc{\'{\i}}a{-}Valls.
\newblock A qos registry for adaptive real-time service-oriented applications.
\newblock In {\em 2011 {IEEE} International Conference on Service-Oriented
  Computing and Applications, {SOCA} 2011, Irvine, CA, USA, December 12-14,
  2011}, pages 1--8, 2011.

\bibitem{6627810}
D.~Bender and K.~Sartipi.
\newblock Hl7 fhir: An agile and restful approach to healthcare information
  exchange.
\newblock In {\em Proceedings of the 26th IEEE International Symposium on
  Computer-Based Medical Systems}, pages 326--331, June 2013.

\bibitem{BersaniHASE18}
Marcello~M. Bersani and Marisol Garc{\'{\i}}a{-}Valls.
\newblock The cost of formal verification in adaptive {CPS.} an example of a
  virtualized server node.
\newblock In {\em 17th {IEEE} International Symposium on High Assurance Systems
  Engineering, {HASE} 2016, Orlando, FL, USA, January 7-9, 2016}, pages 39--46,
  2016.

\bibitem{Bersani18}
Marcello~M. Bersani and Marisol Garc{\'{\i}}a{-}Valls.
\newblock Online verification in cyber-physical systems: Practical bounds for
  meaningful temporal costs.
\newblock {\em Journal of Software: Evolution and Process}, 30(3), 2018.

\bibitem{8363196}
A.~Bestbier and P.~R. Fourie.
\newblock Development of a vital signs monitoring wireless ear probe.
\newblock In {\em 2018 3rd Biennial South African Biomedical Engineering
  Conference (SAIBMEC)}, pages 1--5, April 2018.

\bibitem{mata}
Alberto~Montilla Bravo and Marisol Garc{\'{\i}}a{-}Valls.
\newblock Fipa-based qos negotiator for nomadic agents.
\newblock In {\em Mobile Agents for Telecommunication Applications, 4th
  International Workshop, {MATA} 2002 Barcelona, Spain, October 23-24, 2002,
  Proceedings}, pages 216--226, 2002.

\bibitem{peteradaeurope}
Peter~T. Breuer and Marisol Garc{\'{\i}}a{-}Valls.
\newblock Static deadlock detection in the linux kernel.
\newblock In {\em Reliable Software Technologies - Ada-Europe 2004, 9th
  Ada-Europe International Conference on Reliable Software Technologies, Palma
  de Mallorca, Spain, June 14-18, 2004, Proceedings}, pages 52--64, 2004.

\bibitem{peterspe}
Peter~T. Breuer and Marisol Garc{\'{\i}}a{-}Valls.
\newblock Raiding the noosphere: the open development of networked {RAID}
  support for the linux kernel.
\newblock {\em Softw., Pract. Exper.}, 36(4):365--395, 2006.

\bibitem{demiguel}
M.~A. de~Miguel, J.~F. Ruiz, and M.~Garcia.
\newblock Qos-aware component frameworks.
\newblock In {\em IEEE 2002 Tenth IEEE International Workshop on Quality of
  Service (Cat. No.02EX564)}, pages 161--169, May 2002.

\bibitem{7988050}
A.~J. Dinu, R.~Ganesan, A.~A. Kebede, and B.~Veerasamy.
\newblock Performance analysis and comparison of medical image compression
  techniques.
\newblock In {\em 2016 International Conference on Control, Instrumentation,
  Communication and Computational Technologies (ICCICCT)}, pages 738--745, Dec
  2016.

\bibitem{ekelman1988technological}
Karen~B Ekelman et~al.
\newblock Technological innovation and medical devices.
\newblock 1988.

\bibitem{Fielding:2000:ASD:932295}
Roy~Thomas Fielding.
\newblock {\em Architectural Styles and the Design of Network-based Software
  Architectures}.
\newblock PhD thesis, 2000.
\newblock AAI9980887.

\bibitem{fdainteroperability}
Foof and Drug Administration.
\newblock Medical {D}evice {I}nteroperability.
\newblock \url{https://www.fda.gov/MedicalDevices/DigitalHealth/ucm512245.htm}.
\newblock Accessed: 2018-07-01.

\bibitem{french2012medical}
Elaine French-Mowat and Joanne Burnett.
\newblock How are medical devices regulated in the european union?
\newblock {\em Journal of the Royal Society of Medicine}, 105(1\_suppl):22--28,
  2012.

\bibitem{7325198}
K.~Fysarakis, D.~Mylonakis, C.~Manifavas, and I.~Papaefstathiou.
\newblock Node.dpws: Efficient web services for the internet of things.
\newblock {\em IEEE Software}, 33(3):60--67, May 2016.

\bibitem{isorcyork}
Marisol Garc{\'{\i}}a{-}Valls.
\newblock A proposal for cost-effective server usage in {CPS} in the presence
  of dynamic client requests.
\newblock In {\em 19th {IEEE} International Symposium on Real-Time Distributed
  Computing, {ISORC} 2016, York, United Kingdom, May 17-20, 2016}, pages
  19--26, 2016.

\bibitem{modechanges}
Marisol Garc{\'{\i}}a{-}Valls, Alejandro Alonso, and Juan~Antonio de~la Puente.
\newblock Mode change protocols for predictable contract-based resource
  management in embedded multimedia systems.
\newblock In {\em International Conference on Embedded Software and Systems,
  {ICESS} '09, Hangzhou, Zhejiang, P. R. China, May 25-27, 2009.}, pages
  221--230, 2009.

\bibitem{HolaQOS}
Marisol Garc{\'i}a-Valls, Alejandro Alonso, Jos{\'e} Ruiz, and Angel Groba.
\newblock {\em An Architecture of a Quality of Service Resource Manager
  Middleware for Flexible Embedded Multimedia Systems}, pages 36--55.
\newblock Springer Berlin Heidelberg, Berlin, Heidelberg, 2003.

\bibitem{RPiDDS}
Marisol Garc{\'{\i}}a{-}Valls, Javier Ampuero{-}Calleja, and Luis~Lino
  Ferreira.
\newblock Integration of data distribution service and raspberry pi.
\newblock In {\em Green, Pervasive, and Cloud Computing - 12th International
  Conference, {GPC} 2017, Cetara, Italy, May 11-14, 2017, Proceedings}, pages
  490--504, 2017.

\bibitem{omacy}
Marisol Garc{\'{\i}}a{-}Valls and Roberto Baldoni.
\newblock Adaptive middleware design for {CPS:} considerations on the os,
  resource managers, and the network run-time.
\newblock In {\em Proceedings of the 14th International Workshop on Adaptive
  and Reflective Middleware, ARM@Middleware 2015, Vancouver, BC, Canada,
  December 7-11, 2015}, pages 3:1--3:6, 2015.

\bibitem{FGCSreview}
Marisol Garc{\'{\i}}a{-}Valls, Paolo Bellavista, and Aniruddha~S. Gokhale.
\newblock Reliable software technologies and communication middleware: {A}
  perspective and evolution directions for cyber-physical system, mobility, and
  cloud computing.
\newblock {\em Future Generation Comp. Syst.}, 71:171--176, 2017.

\bibitem{sac2017}
Marisol Garc{\'{\i}}a{-}Valls and Christian Calva{-}Urrego.
\newblock Improving service time with a multicore aware middleware.
\newblock In {\em Proceedings of the Symposium on Applied Computing, {SAC}
  2017, Marrakech, Morocco, April 3-7, 2017}, pages 1548--1553, 2017.

\bibitem{CSIcpsmiddleware}
Marisol Garc{\'{\i}}a{-}Valls, Christian Calva{-}Urrego, Juan~Antonio de~la
  Puente, and Alejandro Alonso.
\newblock Adjusting middleware knobs to assess scalability limits of
  distributed cyber-physical systems.
\newblock {\em Computer Standards {\&} Interfaces}, 51:95--103, 2017.

\bibitem{uc3mupv}
Marisol Garc{\'{\i}}a{-}Valls, Christian Calva{-}Urrego, and Ana
  Garc{\'{\i}}a{-}Fornes.
\newblock Accelerating smart e{H}ealth services execution at the fog computing
  infrastructure.
\newblock {\em Future Generation Comp. Syst.}, --:--, 2018.

\bibitem{JSAReiser}
Marisol Garc{\'{\i}}a{-}Valls, Antonio Casimiro, and Hans~P. Reiser.
\newblock A few open problems and solutions for software technologies for
  dependable distributed systems.
\newblock {\em Journal of Systems Architecture - Embedded Systems Design},
  73:1--5, 2017.

\bibitem{jsareaction2012}
Marisol Garc{\'{\i}}a{-}Valls and Tommaso Cucinotta.
\newblock Real-time and distributed computing in emerging applications.
  foreword by the general chairs of reaction 2012.
\newblock {\em Journal of Systems Architecture - Embedded Systems Design},
  61(5-6):267--268, 2015.

\bibitem{DDSFace}
Marisol Garc{\'{\i}}a{-}Valls, Jorge Dom{\'{\i}}nguez{-}Poblete, Imad~Eddine
  Touahria, and Chenyang Lu.
\newblock Integration of data distribution service and distributed partitioned
  systems.
\newblock {\em Journal of Systems Architecture - Embedded Systems Design},
  83:23--31, 2018.

\bibitem{uc3muma}
Marisol Garc{\'{\i}}a{-}Valls, Daniel Garrido, and Manuel D{\'{\i}}az.
\newblock Impact of middleware design on the communication performance.
\newblock In {\em Green, Pervasive, and Cloud Computing - 12th International
  Conference, {GPC} 2017, Cetara, Italy, May 11-14, 2017, Proceedings}, pages
  505--519, 2017.

\bibitem{sigbediland}
Marisol Garc{\'{\i}}a{-}Valls, Natividad Herrasti, Christophe Jouvray, and
  Aintzane Armentia.
\newblock Flexible and timely on-line integration of medical services using
  iland middleware.
\newblock {\em {SIGBED} Review}, 14(2):53--60, 2017.

\bibitem{adaeurope12}
Marisol Garc{\'{\i}}a{-}Valls and Felipe Ib{\'{a}}{\~{n}}ez{-}V{\'{a}}zquez.
\newblock Integrating middleware for timely reconfiguration of distributed soft
  real-time systems with ada {DSA}.
\newblock In {\em Reliable Software Technologies - Ada-Europe 2012 - 17th
  Ada-Europe International Conference on Reliable Software Technologies,
  Stockholm, Sweden, June 11-15, 2012. Proceedings}, pages 35--48, 2012.

\bibitem{ilandTII}
Marisol Garc{\'{\i}}a{-}Valls, Iago~Rodr{\'{\i}}guez Lopez, and Laura
  Fern{\'{a}}ndez{-}Villar.
\newblock iland: An enhanced middleware for real-time reconfiguration of
  service oriented distributed real-time systems.
\newblock {\em {IEEE} Trans. Industrial Informatics}, 9(1):228--236, 2013.

\bibitem{aina}
Marisol Garc{\'{\i}}a{-}Valls, Alejandro~Alonso Mu{\~{n}}oz, and Juan~Antonio
  de~la Puente.
\newblock Time-predictable reconfiguration with contract-based resource
  management.
\newblock In {\em 23rd International Conference on Advanced Information
  Networking and Applications, {AINA} 2009, Workshops Proceedings, Bradford,
  United Kingdom, May 26-29, 2009}, pages 494--499, 2009.

\bibitem{Compsac2014}
Marisol Garc{\'{\i}}a{-}Valls, Diego Perez{-}Palacin, and Raffaela Mirandola.
\newblock Time-sensitive adaptation in {CPS} through run-time configuration
  generation and verification.
\newblock In {\em {IEEE} 38th Annual Computer Software and Applications
  Conference, {COMPSAC} 2014, Vasteras, Sweden, July 21-25, 2014}, pages
  332--337, 2014.

\bibitem{JSSPoliMi}
Marisol Garc{\'{\i}}a{-}Valls, Diego Perez{-}Palacin, and Raffaela Mirandola.
\newblock Pragmatic cyber physical systems design based on parametric models.
\newblock {\em Journal of Systems and Software}, 144:559--572, 2018.

\bibitem{sensorsice}
Marisol Garc{\'{\i}}a{-}Valls and Imad~Eddine Touahria.
\newblock On line service composition in the integrated clinical environment
  for ehealth and medical systems.
\newblock {\em Sensors}, 17(6):1333, 2017.

\bibitem{JSASI2017}
Marisol García-Valls, António Casimiro, and Hans~P. Reiser.
\newblock A few open problems and solutions for software technologies for
  dependable distributed systems.
\newblock {\em Journal of Systems Architecture}, 73:1 -- 5, 2017.
\newblock Special Issue on Reliable Software Technologies for Dependable
  Distributed Systems.

\bibitem{goldman2008medical}
Julian~M Goldman.
\newblock Medical devices and medical systems essential safety requirements for
  equipment comprising the patient-centric integrated clinical environment
  ({ICE})-{P}art 1: General requirements and conceptual model.
\newblock {\em ASTM International}, 2008.

\bibitem{groba}
Angel~M. Groba, Alejandro Alonso, Jos{\'{e}}~A. Rodr{\'{\i}}guez, and Marisol
  Garc{\'{\i}}a{-}Valls.
\newblock Response time of streaming chains: Analysis and results.
\newblock In {\em 14th Euromicro Conference on Real-Time Systems {(ECRTS}
  2002), 19-21 June 2002, Vienna, Austria, Proceedings}, page 182, 2002.

\bibitem{8031128}
N.~Hong, K.~Wang, L.~Yao, and G.~Jiang.
\newblock Visual fhir: An interactive browser to navigate hl7 fhir
  specification.
\newblock In {\em 2017 IEEE International Conference on Healthcare Informatics
  (ICHI)}, pages 26--30, Aug 2017.

\bibitem{7390446}
M.~Kasparick, S.~Schlichting, F.~Golatowski, and D.~Timmermann.
\newblock Medical dpws: New ieee 11073 standard for safe and interoperable
  medical device communication.
\newblock In {\em 2015 IEEE Conference on Standards for Communications and
  Networking (CSCN)}, pages 212--217, Oct 2015.

\bibitem{5070972}
A.~King, S.~Procter, D.~Andresen, J.~Hatcliff, S.~Warren, W.~Spees, R.~Jetley,
  P.~Jones, and S.~Weininger.
\newblock An open test bed for medical device integration and coordination.
\newblock In {\em 2009 31st International Conference on Software Engineering -
  Companion Volume}, pages 141--151, May 2009.

\bibitem{7814546}
M.~Lepmets, T.~McBride, and F.~McCaffery.
\newblock Towards safer medical device software systems: Industry-wide learning
  from failures and the use of safety-cases to support process compliance.
\newblock In {\em 2016 10th International Conference on the Quality of
  Information and Communications Technology (QUATIC)}, pages 193--198, Sept
  2016.

\bibitem{7494019}
Hai-Long Li, Zhi-Bin Duan, Jin-Zhong Cui, and Zhen-Wei Chen.
\newblock A design of general medical data adapter based on iso/ieee 11073
  standards.
\newblock In {\em 2015 12th International Computer Conference on Wavelet Active
  Media Technology and Information Processing (ICCWAMTIP)}, pages 404--407, Dec
  2015.

\bibitem{7336385}
J.~Lu, W.~Hu, M.~Song, X.~Zhan, and X.~Liu.
\newblock Mobile medical service system based on portable devices.
\newblock In {\em 2015 IEEE 17th International Conference on High Performance
  Computing and Communications, 2015 IEEE 7th International Symposium on
  Cyberspace Safety and Security, and 2015 IEEE 12th International Conference
  on Embedded Software and Systems}, pages 1530--1535, Aug 2015.

\bibitem{5974909}
X.~Lu, Y.~Gu, J.~Zhao, N.~Yu, and W.~Jia.
\newblock Research and implementation of medical information format conversion
  based on hl7 version 2.x.
\newblock In {\em 2011 International Conference on Computer Science and Service
  System (CSSS)}, pages 2440--2443, June 2011.

\bibitem{6334808}
M.~Mchugh, F.~Mccaffery, and V.~Casey.
\newblock Software process improvement to assist medical device software
  development organisations to comply with the amendments to the medical device
  directive.
\newblock {\em IET Software}, 6(5):431--437, October 2012.

\bibitem{6152915}
J.~Nam, W.~Seo, J.~Bae, and Y.~Cho.
\newblock Design and development of a u-health system based on the iso/ieee
  11073 phd standards.
\newblock In {\em The 17th Asia Pacific Conference on Communications}, pages
  789--793, Oct 2011.

\bibitem{5687707}
R.~Noumeir and J.~F. Pambrun.
\newblock Hands-on approach for teaching hl7 version 3.
\newblock In {\em Proceedings of the 10th IEEE International Conference on
  Information Technology and Applications in Biomedicine}, pages 1--4, Nov
  2010.

\bibitem{DDS}
OMG.
\newblock Data distribution software, v1.4.
\newblock https://www.omg.org/spec/DDS/, 2015.

\bibitem{world2003diet}
World~Health Organization.
\newblock {\em Diet, nutrition, and the prevention of chronic diseases: report
  of a joint WHO/FAO expert consultation}, volume 916.
\newblock World Health Organization, 2003.

\bibitem{ambient}
Clara Otero-P\'{e}rez, Liesbeth Steffens, Peter van~der Stock, Sjir van Loo,
  Alejandro Alonso, Jos\'{e} Ruiz, Reinder Brill, and Marisol Garc\'{i}a-Valls.
\newblock Qos-based resource management for ambient intelligence.
\newblock 2003.

\bibitem{8336776}
V.~Randazzo, E.~Pasero, and S.~Navaretti.
\newblock Vital-ecg: A portable wearable hospital.
\newblock In {\em 2018 IEEE Sensors Applications Symposium (SAS)}, pages 1--6,
  March 2018.

\bibitem{adaeurope11}
Iago Rodr{\'{\i}}guez{-}L{\'{o}}pez and Marisol Garc{\'{\i}}a{-}Valls.
\newblock Architecting a common bridge abstraction over different middleware
  paradigms.
\newblock In {\em Reliable Software Technologies - Ada-Europe 2011 - 16th
  Ada-Europe International Conference on Reliable Software Technologies,
  Edinburgh, UK, June 20-24, 2011. Proceedings}, pages 132--146, 2011.

\bibitem{sigbedamqp}
Paloma Rubio{-}Conde, Diego Villar{\'{a}}n{-}Molina, and Marisol
  Garc{\'{\i}}a{-}Valls.
\newblock Measuring performance of middleware technologies for medical systems:
  Ice vs {AMQP}.
\newblock {\em {SIGBED} Review}, 14(2):8--14, 2017.

\bibitem{7377675}
M.~I. Sabar, P.~M. Jayaweera, and E.~A. T.~A. Edirisuriya.
\newblock International interoperability through unified universal hl7 v3 green
  messaging.
\newblock In {\em 2015 Fifteenth International Conference on Advances in ICT
  for Emerging Regions (ICTer)}, pages 112--118, Aug 2015.

\bibitem{7318708}
J.~Schlamelcher, M.~Onken, M.~Eichelberg, and A.~Hein.
\newblock Dynamic dicom configuration in a service-oriented medical device
  architecture.
\newblock In {\em 2015 37th Annual International Conference of the IEEE
  Engineering in Medicine and Biology Society (EMBC)}, pages 1717--1720, Aug
  2015.

\bibitem{staff2010iso}
ISO Staff.
\newblock Iso/ieee 11073-20601: Health informatics--personal health device
  communication; part 20601 application profile-optimized exchange protocol.
\newblock {\em Ginebra, Suiza}, 2010.

\bibitem{8029619}
I.~E. Touahria, M.~Garc\'{i}a-Valls, and A.~Khababa.
\newblock An ice compliant component model for medical systems development.
\newblock In {\em 2017 IEEE 41st Annual Computer Software and Applications
  Conference (COMPSAC)}, volume~1, pages 278--287, July 2017.

\end{thebibliography}
\end{document}